\documentclass[preprint2]{aastex631}
\begin{document}
\journalinfo{Accepted for Publication (MNRAS)}
\shorttitle{Concerning S$H_0$ES $W_{VI}$ data (NGC4258)}
\title{Concerning S$H_0$ES Data: Discrepant $W_{0,VI}$ Absolute Magnitudes for Cepheids in the Keystone Galaxy NGC4258}

\author[0000-0001-8803-3840]{Daniel Majaess}
\affiliation{Mount Saint Vincent University, Halifax, Canada}
\email{daniel.majaess@msvu.ca}

\begin{abstract}
S$H_0$ES $VI$-band photometry for classical Cepheids in the keystone galaxy NGC4258 yield discrepant absolute magnitudes. Specifically, the 2016 and 2022 published S$H_0$ES Cepheid data for NGC4258 exhibit a substantial offset of $\Delta W_{0,VI}\simeq0^{\rm m}.3$. That adds to a suite of existing concerns associated with the S$H_0$ES analysis, which in sum imply that their relatively non-changing Hubble constant for nearly twenty years warrants scrutiny. 
\end{abstract}


\section{Introduction}
\label{sec:intro}
The S$H_0$ES results are leveraged to advocate that classical Cepheid distances yield an $H_0$ that is offset from the Planck CMB result \citep{rie22}.  There are indeed concerns with $\Lambda$CDM, however, there exist errors and anomalies within S$H_0$ES data that provide sufficient pause when considering their conclusions, despite potentially being fortuitously correct.  That includes inconsistent S$H_0$ES photometry for Cepheids in remote galaxies with inhomogeneous crowding and surface brightness profiles \citep[e.g.,][]{efs20,fm23b}, a suite of contested Leavitt Law parameters such as slope, extinction law, metallicity \citep[e.g.,][]{mf23}, and changes in maser and Cepheid distances to the keystone galaxy NGC4258 \citep[e.g.,][]{maj10,maj24}.  Yet, the S$H_0$ES $H_0 \simeq 73$ km/s/Mpc remained comparatively unaltered across approximately two decades.  A subset of such disconcerting issues are highlighted below, and the new analysis is presented in \S \ref{sec:analysis}.

\citet{maj10} argued that the $W_{VI}$ slope of the Leavitt Law for near-solar Cepheids determined by \citet[][S$H_0$ES]{rie09b} exhibited a marked offset ($\alpha=-2.98\pm0.07$) from the consensus result \citep[$\alpha\simeq -3.3$, see also][]{maj11,rie22}.  \citet{maj10} further remarked that a shallower $W_{VI}$ slope may be indicative of incorrect photometric standardization or decontamination procedures (crowding/blending), and that certain S$H_0$ES V-I colors may be too blue \citep[see also][]{efs20}.  \citet{fm23b} stressed that, \textit{``A simple comparison of the distance moduli tabulated in Riess et al.~(2016, 2022) reveals an overall difference of $-0^{\rm m}.123$ ... That corresponds to a 6\% shift in $H_0$."} 

\citet{efs20} determined the following regarding the $W_{H-VI}$ data of \citet[][S$H_0$ES]{rie16}, \textit{``The LMC distance\footnote{\citet{pie19}} together with the S$H_0$ES Cepheids is placing NGC4258 at a distance of $6.98$ Mpc if metallicity effects are ignored, whereas the maser distance is $7.58$ Mpc."}  \citet{maj24} conveyed that the \citet{mac06}, \citet[][S$H_0$ES]{hof16}, and \citet[][S$H_0$ES]{yua22} $W_{VI}$ Cepheid datasets for NGC4258 are discordant, both \textit{vis \`a vis} the mean distance and the impact of chemical composition on Cepheid distances.  \citet{mac06} and \citet[][S$H_0$ES]{hof16}\footnote{See the broader companion study of \citet[][S$H_0$ES]{rie16}.} favored a relatively stronger dependence of the $W_{VI}$ Leavitt Law zeropoint on metallicity, while the \citet[][S$H_0$ES]{yua22} data suggest otherwise \citep[see Fig.~3 in][]{maj24}. \citet{uda01}, \citet{maj11}, \citet{wie17}, and \citet{mf23} concluded that $W_{VI}$ functions are relatively insensitive to metallicity,\footnote{\citet{bre22} provide a rebuttal that the reader can consider in their \S 5.5, and see also their Table~1. Disagreements likewise exist from the important perspective of modelling \citep{bon08,and16}.} whereas \citet[][S$H_0$ES]{rb23} advocate for a larger zeropoint dependence by comparison (e.g., $\gamma= -0.22\pm0.04$ mag/dex). For example, \citet{mf23} relayed TRGB-Cepheid distances (their Fig.~1) which contest that \citet[][S$H_0$ES]{rb23} finding.  The reader can examine Fig.~6 in \citet[][S$H_0$ES]{yua22} and assess whether a constant (indicating a null-dependence) represents their latest NGC4258 analysis rather than the fits they overlaid, and pair that with an inspection of the extended metallicity baseline present in Fig.~1 of \citet{mf23}.  \citet[][S$H_0$ES]{yua22} determined that $\gamma$ is $-0.07\pm0.21$ mag/dex.  Moreover, \citet{maj24} added there is an alarming $I$-band (F814W) and $W_{VI}$ discrepancy between \citet[][S$H_0$ES]{hof16} and \citet[][S$H_0$ES]{yua22}, which is characterized by a considerable mean difference ($\gtrsim 0^{\rm m}.15$).   That could stem from inhomogeneous photometry or crowding corrections, and \citet[][S$H_0$ES]{yua22} stated, \textit{``many past works have not fully incorporated individual Cepheid crowding corrections (e.g., Macri et al.~2006, Hoffmann et al.~2016) as we have here, which will make the Cepheids fainter."}  Cepheids at smaller galactocentric radii can feature enhanced chemical abundances and be projected upon an increased stellar density and high surface brightness background.  That degeneracy has compromised certain determinations of the impact of metallicity on Cepheid distances \citep[\S 5 of][and references therein]{Mac01,maj11}, and photometric contamination has propagated a systematic uncertainty into $H_0$ determinations and the cosmic distance scale \citep[e.g.,][]{su99,moc04,maj12}.

\citet{maj20} readily identified blended LMC RR Lyrae variables, whereas brighter Cepheids were more challenging to differentiate in that respect, and thus it was emphasized that the S$H_0$ES approach to decontamination of more remote Cepheids must be independently verified, since they applied significant crowding corrections \citep[e.g., $0^{\rm m}.22$, Table~2 in][]{rie11}.  Independently, \citet{fm23} relayed an analysis by I.~Jang regarding the \citet[][S$H_0$ES]{rie12} crowding data, noting, \textit{``With a median correction of $\sim0^{\rm m}.25$, which corresponds to a 10\% difference in $H_0$ of $>7$ km s$^{-1}$ Mpc$^{-1}$ ... there remains the potential for a hidden systematic effect that may be difficult to identify and account for."}  

\begin{deluxetable*}{lc}
\tablecaption{S$H_0$ES data for classical Cepheids in the keystone galaxy NGC4258 provide discordant absolute magnitudes ($W_{0,VI}$).\label{table}}
\tablehead{\colhead{$W_{VI}$ dataset} & \colhead{$\beta$ (Eqn.~\ref{eqn:abs})}}
\startdata
\citet[][S$H_0$ES]{hof16} & $-2.97\pm0.04$  \\
\citet[][S$H_0$ES]{yua22} & $-2.66\pm0.05$  \\
\enddata
\end{deluxetable*}

In this study, additional points of concern are highlighted regarding S$H_0$ES, with a focus on the incompatible absolute magnitudes ($W_{0,VI}$) implied for NGC4258 Cepheids across the following studies: \citet[][S$H_0$ES]{hof16} and \citet[][S$H_0$ES]{yua22}.

\section{Analysis}
\label{sec:analysis}
The \citet[][S$H_0$ES]{hof16} and \citet[][S$H_0$ES]{yua22} datasets are compared with respect to the absolute magnitudes implied by the Araucaria distance to the LMC \citep{pie19}, and the maser distance to NGC4258 \citep{rei19}.  Those are anchor points adopted by S$H_0$ES.  However, such oft-cited distances may be incorrect, and there are LMC estimates with relatively low \textit{cited} uncertainties \citep[e.g.,][and references therein, and the reader should likewise weigh the conclusions of \citealt{sch08}]{ste20}.  

First, the apparent Wesenheit ($W_{VI}$) magnitude is given by:
\begin{equation}
W_{VI}=F814W-1.45(F555W-F814W)
\end{equation}
The color coefficient represents the extinction law adopted by OGLE and \citet[][S$H_0$ES]{hof16}.
The form for the absolute $W_{0,VI}$ magnitude is:
\begin{equation}
\label{eqn:abs}
W_{0,VI}=\alpha \log{P} + \beta
\end{equation}
The distance modulus follows as:
\begin{equation}
W_{VI}-W_{0,VI} = \mu_0
\end{equation}
The coefficient and zeropoint can be determined by combining the expressions:
\begin{equation}
F814W-1.45(F555W-F814W) = \alpha \log{P} + \beta + \mu_0
\end{equation}

The \citet{rie19} data for LMC Cepheids, in tandem with their adopted $\mu_0=18.477\pm0.026$ \citep{pie19}, yield the following results via minimization:
\begin{equation}
\beta=-2.65\pm0.04 , \alpha=-3.34\pm 0.03
\label{eqn:lmc}
\end{equation}
Where $\beta$ is the zeropoint of the absolute Wesenheit magnitude, and which shall be compared to that inferred from NGC4258 data.  Note the slope $\alpha$ is comparable to that determined by \citet{maj10} and \citet{maj11} for Local Group Cepheids, and across a sizable metallicity baseline.  That starkly contrasts earlier findings by \citet[][S$H_0$ES]{rie09b}. The \citet[][S$H_0$ES]{hof16} data for NGC4258 Cepheids, in tandem with the \citet{rei19} maser distance ($\mu_0=29.397\pm0.032$), yield the following results via minimization:
\begin{equation}
\beta=-3.26\pm0.05, \alpha=-3.09\pm0.03
\end{equation}
Critically, those findings vastly differ from Eqn.~\ref{eqn:lmc}.  Hereon, the LMC slope will be adopted following  the \citet{maj10} and \citet{maj11} results, whose analysis relied in part on Araucaria, OGLE, and \citet{ben07} data. Redoing the analysis with the LMC slope and NGC4258 maser distance yields:
\begin{equation}
\beta=-2.97 \pm 0.04
\end{equation}
Note the substantial difference ($\simeq 0^{\rm m}.3$) relative to the LMC determined absolute magnitude (Eqn.~\ref{eqn:lmc}), or that of \citet[][S$H_0$ES]{yua22} (Eqn.~\ref{eqn:yuan}). Even if the \citet[][S$H_0$ES]{rb23} metallicity effect was adopted ($\gamma =0.22\pm0.04$ mag/dex), despite arguments to the contrary \citep[e.g.,][]{mf23}: the ensuing $<0^{\rm m}.1$ correction is far from reconciling the absolute magnitudes.  \citet{efs20} examined the $W_{H-VI}$ function and discovered a $0.177\pm0^{\rm m}.051$ discrepancy when comparing distance moduli established for NGC4258 using the \citet[][S$H_0$ES]{rie16} observations, the \citet{rie19} LMC observations, and anchor points for those galaxies \citep{rei19,pie19}.

The more recent $W_{VI}$ S$H_0$ES analysis by \citet[][S$H_0$ES]{yua22} of NGC4258 Cepheids essentially eliminates the aforementioned $\simeq 0^{\rm m}.3$ deviation tied to the earlier S$H_0$ES photometry \citep{hof16}, and the minimization procedure yielded:
\begin{equation}
\beta=-2.66 \pm 0.05
\label{eqn:yuan}
\end{equation}
Perhaps the revised S$H_0$ES approach was motivated in part by criticisms concerning photometry, crowding, blending, and the resulting offsets \citep[e.g., Eqn.~3.5 in][]{efs20}, and possibly by the \citet{yua20} recognition that Cepheids in high surface brightness regions of the Seyfert 1 galaxy NGC4151 exhibited a systematic shift (their Fig.~9, right panel).  Nevertheless, the S$H_0$ES photometry for the keystone galaxy NGC4258 is discrepant across time.  Admittedly, both the maser and Cepheid distances to NGC4258 feature a discouraging history, with early estimates of the former \citep[$6.4\pm0.9$ and $7.2\pm0.3$ Mpc,][]{miy95,her99} being extensively nearer than the \citet{rei19} result of $7.576\pm0.082\pm0.076$ Mpc. \citet{maj24} conveyed that the suite of $W_{VI}$ Cepheid datasets for NGC4258 exhibited divergent results \citep{mao99,new01,mac06,fau15,hof16,yua22}.  

The \citet[][S$H_0$ES]{hof16} and \citet[][S$H_0$ES]{yua22} data for NGC4258 provide irreconcilable absolute magnitudes.  That result likewise holds when adopting an $\alpha(\log{P}-1)$ framework for Eqn.~\ref{eqn:abs}.

\section{Conclusion}
$W_{VI}$ photometry of Cepheids in NGC4258 is inconsistent between \citet[][S$H_0$ES]{hof16} and \citet[][S$H_0$ES]{yua22}, in terms of the mean implied distance, the impact of metallicity, and with respect to the absolute magnitude constrained by LMC and NGC4258 (M106) distances adopted by the S$H_0$ES team (Table~\ref{table}).  The latter $\simeq 0^{\rm m}.3$ discrepancy is too large to be explained by the contested metallicity effect proposed by S$H_0$ES \citep[][see also the replies within \citealt{efs20}]{rb23}.  

More broadly, there are numerous inconsistencies endemic to S$H_0$ES data over time, as outlined previously \citep[e.g.,][]{fm23b,maj24}. The following overarching conclusions regarding Cepheids are worth re-emphasizing, namely: 

(1) The dawn of precision cosmology seemingly occurs in an era where a lack of agreement exists concerning fundamentals associated with the Leavitt Law, owing in part to degeneracies (e.g., crowding, blending, metallicity, and extinction law).  Additional research on the latter topic tied to variations in the extinction law is of particular interest \citep[e.g.,][]{tur13,fau15}. 

(2) A $W_{VI}$ metallicity effect is not a chief source of uncertainty associated with Cepheid distances or the establishment of $H_0$, but rather it is the challenging task of obtaining precise, commonly standardized, multiepoch, multiband, comparatively uncontaminated extragalactic Cepheid photometry. 

(3) A consensus framework to assess photometric contamination should be pursued, in concert with an elaborate characterization of the difference in approach to crowding adopted by \citet[][S$H_0$ES]{yua22} relative to previous work \citep{rie09b,rie11,rie16,hof16}, especially given Table~\ref{table} and the passage on crowding from \citet[][S$H_0$ES]{yua22} restated verbatim in \S \ref{sec:intro}, and owing to anomalies described in the literature \citep[e.g.,][]{maj10,efs20,fm23b,maj24}.  Indeed, the entire suite of raw HST images must be reassessed by independent researchers to benchmark S$H_0$ES assertions throughout the project's history, since confirmation bias can unwittingly impact conclusions.  Examining the veracity of prior S$H_0$ES findings provides guidance on whether the cited Hubble constant and uncertainties are reliable.  Importantly, the full Cepheid (unculled) datasets should be published \citep{efs20}.    

Consequently, TRGB and JAGB distances in \textit{non-crowded} regions are desirable \citep[e.g.,][]{mf23c,fm23}.

\begin{acknowledgments}
\textbf{Acknowledgments}: this research relies on the efforts of the following initiatives: CDS, NASA ADS, arXiv, OGLE, Araucaria, S$H_0$ES, (C)CHP.
\end{acknowledgments}

\bibliography{article}{}
\bibliographystyle{aasjournal}



\end{document}